\RequirePackage{fix-cm}
\documentclass{svjour3}                     
\smartqed  
\usepackage{graphicx}
\usepackage{amsmath}
\usepackage{amssymb}
\usepackage{booktabs}
\usepackage{multirow}
\begin{document}

\title{Refractive-index sensing with ultra-thin plasmonic nanotubes}

\author{S.~Raza         \and
        G.~Toscano     \and
        A.-P.~Jauho    \and
        N.~A.~Mortensen        \and
        M.~Wubs
}

\institute{S. Raza
                \at{Department of Photonics Engineering, Technical University of Denmark, DK-2800 Kgs. Lyngby, Denmark
                \and Center for Electron Nanoscopy, Technical University of Denmark, DK-2800 Kgs. Lyngby, Denmark}
            \and
           G. Toscano \and N. A. Mortensen \and M. Wubs
                \at Department of Photonics Engineering, Technical University of Denmark, DK-2800 Kgs. Lyngby, Denmark \\
                \email{mwubs@fotonik.dtu.dk}
           \and
           A.-P. Jauho
                \at Department of Micro- and Nanotechnology, Technical University of Denmark, DK-2800 Kgs. Lyngby, Denmark
}


\maketitle

\begin{abstract}
We study the refractive-index sensing properties of plasmonic nanotubes with a dielectric core and ultra-thin metal shell. The few-nm thin metal shell is described by both the usual Drude model and the nonlocal hydrodynamic model to investigate the effects of nonlocality. We derive an analytical expression for the extinction cross section and show how sensing of the refractive index of the surrounding medium and the figure-of-merit are affected by the shape and size of the nanotubes. Comparison with other localized surface plasmon resonance sensors reveals that the nanotube exhibits superior sensitivity and comparable figure-of-merit.
\end{abstract}

\section{Introduction}
It is well known that metallic nanoparticles can sustain localized surface plasmon (LSP) oscillations, whose resonance frequencies in the quasi-static limit depend solely on the geometry of the nanoparticle, the permittivity of the metal and the surrounding permittivity. The dependency of the LSP resonance (LSPR) on the surrounding medium makes metallic particles extremely good sensors, progressing towards the detection of single molecules \cite{Anker:2008a}. However, the weak effect of retardation on the LSP resonance in nanosized metal particles leaves only one parameter to truly engineer: the geometry. By modifying the structure of the metal nanoparticle to have a dielectric core with a metal shell, an increased tunability is achieved due to the plasmon hybridization of the inner and outer surfaces of the metal \cite{Prodan:2003a}. Especially the spherical core-shell structure has received a considerable amount of attention in recent years \cite{Brongersma:2003a,Raschke:2004a,Nehl:2004a,Tam:2004a} due to its excellent and tunable sensing properties, which show great promise in biological studies such as cancer therapy \cite{Bardhan:2011a}. The plasmon hybridization allows one to position the LSP resonance of the nanoshell as desired by simply varying the core size $r_1$ and/or outer radius $r_2$ appropriately~\cite{Prodan:2004a}.

The hybridization of the inner and outer surface plasmons increases when the metal shell becomes thinner \cite{Prodan:2004a}, which gives rise to significantly altered LSP resonances compared to usual homogeneous metal nanoparticles. Studies of the hybridization between two spherical \cite{David:2011a} or cylindrical \cite{Toscano:2012a} metal nanoparticles in few-nm proximity reveal that effects of nonlocal response increase with increasing hybridization. Furthermore, nanosized metal particles \cite{Boardman:1977a,Dasgupta:1981a,Ruppin:1973a,Raza:2011a} and metal films \cite{Jones:1969a} are also strongly affected by nonlocal effects. The core-shell particle thus calls for a nonlocal description, since it features an ultra-thin metallic shell with resulting strong plasmon hybridization.

The use of arrays of nanotubes with high aspect ratio for biosensing \cite{McPhillips:2010a} and hydrogen sensing \cite{Lim:2012a} has yielded impressive results, yet only few theoretical studies have been performed on the nanotube \cite{Schroter:2001a,Zhu:2011a}. Schr\"{o}ter \textit{et al.} investigate the plasmonic modes and dispersion relations of the nanotube \cite{Schroter:2001a}, while Zhu \textit{et al.} perform calculations using the discrete dipole approximation to discuss the changes of the resonance wavelength of the nanotube due to variations of the aspect ratio \cite{Zhu:2011a}. Thus, to our knowledge no systematic study has yet been performed that addresses which parameters determine the LSPR refractive-index sensitivity of a nanotube-based sensor. In this paper, we fill this gap with a systematic study of the sensing and scattering properties of a single infinitely long cylindrical core-shell nanowire (see inset of Fig. \ref{fig:fig1}), which is a good description of dilute arrays of non-interacting nanotubes with high aspect ratio. On the basis of this study, we propose how to optimize a nanotube-based sensor to achieve the utmost sensitivity for the refractive-index sensing of both gases and liquids.

The outline of this paper is as follows. In Sec.~\ref{sec:theory} we discuss the physical principles of local and nonlocal response, and introduce the sensitivity and figure-of-merit (FOM) as quantitative measures of the performance of a LSPR-based sensor. Section~\ref{sec:results} is dedicated to the study of a nanotube with a silica core and gold shell. We determine the dependency of the sensitivity and FOM on the shape and size of the nanotube, using both local and nonlocal theory to model the response of the gold shell. Our conclusions and outlook on nanotube-based sensors is given in Sec.~\ref{sec:con}, and details on the analytical calculations in the Appendix.

\section{Theory} \label{sec:theory}
The ability of LSPR-based sensors to detect changes in the refractive index of their surrounding medium is usually quantified by the sensitivity and FOM \cite{Anker:2008a}. The sensitivity $\partial\lambda/\partial\text{n}_\text{b}$ is determined as the shift in wavelength of the considered LSP resonance in the extinction spectrum of the sensor, when varying the background refractive index $n_\text{b}=\sqrt{\epsilon_\text{b}}$, while the FOM is given as
\begin{equation}
    \text{FOM}=\frac{|\partial\lambda/\partial\text{n}_\text{b}|}{\Delta \lambda} \label{eq:FOM}
\end{equation}
where $\Delta \lambda$ is the resonance linewidth, calculated as the FWHM of the considered LSP resonance in the extinction spectrum. Thus, to determine the performance of the nanotube as a LSPR sensor, we must calculate its extinction cross section, as this quantifies the extinction spectrum and therefore allows us to determine the sensitivity and FOM.

Predictions for the extinction cross section depend on how the optical response of electrons in the metal is modeled. The common approach to describe the response of metals is by making the local approximation which assumes that the response field at a certain position is proportional to the driving field at that position, with the proportionality function being a position- and frequency-dependent dielectric function. This approach has the rather unphysical consequence that all surface charges reside on an infinitely thin layer on the boundaries of the metal, thereby neglecting the actual extent (or wave nature) of the electrons. While the local approximation is justified as long as the metal boundaries are far apart such that the interaction between electrons due to their extent can be neglected (i.e. large metallic structures), it can not be safely assumed for nanosized metal particles where the wavelength of the electron becomes comparable in size to the metal particle. Describing the metal using the semiclassical hydrodynamic Drude model \cite{Raza:2011a}, we relax the local approximation by allowing the existence of local inhomogeneity in the density of the electron gas, which gives rise to pressure waves. The electron-gas pressure waves provide a means to transport energy in the metal in addition to the electromagnetic waves, which gives rise to nonlocal response: the response of the metal at a certain spatial point can depend on the driving field at other nearby points (on the length scale of the Fermi wavelength) in the metal.

In the Appendix, we provide an analytical expression for the extinction cross section in the cases of both nonlocal and local response, for a normally incident TM-polarized wave, see the inset of Fig. \ref{fig:fig1}. We have checked the analytical expression with our numerical implementation of the hydrodynamic Drude model \cite{Toscano:2012a}, which showed perfect agreement (not shown in this paper).

\section{Results and discussion} \label{sec:results}
\begin{figure*}[!tb]
\center
\includegraphics[width=0.9\textwidth]{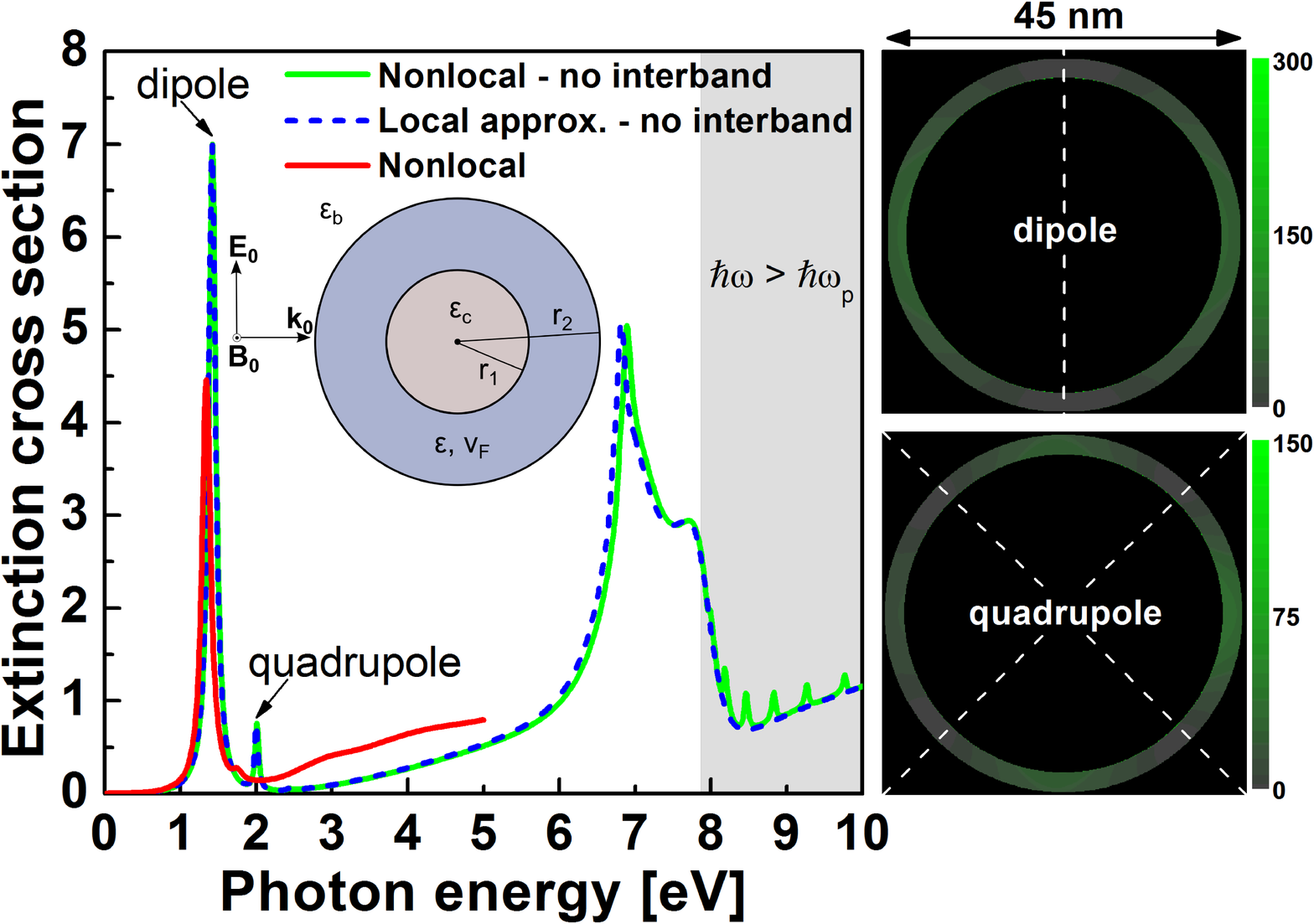}
\caption{Extinction cross sections as a function of incident photon energy for TM-polarized light normally incident on a $(r_1,r_2)=(40\,\text{nm},45\,\text{nm})$ silica-Au cylinder in vacuum. The three curves correspond to the nonlocal and local models without interband transitions (solid green and dashed blue curves, respectively), and the nonlocal model with interband transitions (solid red curve). Free-electron parameters for Au as in Ref. \cite{Rakic:1998a}: $\hbar \omega_{\rm p} = 7.872\,{\rm eV}$, $\hbar\gamma = 0.0530\,{\rm eV}$, and $v_{\rm F} = 1.39 \times 10^{6}\,{\rm m/s}$. Interband parameters for Au are also as in Ref. \cite{Rakic:1998a} and valid up to $5\,{\rm eV}$.
The panel on the right shows the normalized intensity distributions $|\mathbf E|^2/|\mathbf{E_0}|^2$ in the nonlocal model without interband transitions at the dipole and quadrupole resonance frequencies. Here, $\mathbf{E_0}$ is the incident electric field.
 Inset: Schematic diagram of core-shell structure with relevant parameters.}
\label{fig:fig1}
\end{figure*}

We consider the specific core-shell structure, where the core is silica ($\text{SiO}_2$) with dielectric constant $\epsilon_\text{c}=1.5^2$ and the shell is gold (Au) modeled with the data by Raki\'{c} \textit{et al.} \cite{Rakic:1998a}. To clearly show the difference in extinction cross section in local and nonlocal response, we start by examining the case where interband effects in Au are neglected. Figure \ref{fig:fig1} depicts the extinction cross section for a $(r_1,r_2)=(40\,\text{nm},45\,\text{nm})$ silica-Au cylinder in vacuum comparing the local and nonlocal model. The local approximation shows three distinct peaks, two at low frequencies (dipole and quadrupole peaks) and one at a high frequency (near 7\,eV). These are due to the interaction between the localized plasmons at the inner and outer surface of the nanoshell, or equivalently, the interaction between a cavity mode and a cylinder mode \cite{Prodan:2004a}. The nonlocal description allows the same classification of peaks as the local approximation \cite{Raza:2011a,Toscano:2012a}, although the high-frequency peak is blueshifted compared to the local model. Since sensing depends on peak shifts, it is important to take possible nonlocal blueshifts into account. However, the low-frequency resonances show no noticeable blueshift, because the strength of the nonlocal blueshift does not only increase with decreasing thickness of the metal layer \cite{Raza:2011a,David:2011a,Jones:1969a,Ruppin:1973a} but it also depends on the frequency, with a decreasing blueshift for lower frequencies. Thus we find that there is an intricate interplay between plasmon hybridization and nonlocal response: Since a thinner metal shell produces stronger plasmon hybridization, the dipole and quadrupole peaks are pushed to such low frequencies that the nonlocal blueshift effect due to nanosized metallic features is counteracted by the low frequency of the resonances.

The panel on the right of Fig. \ref{fig:fig1} shows the nonlocal normalized intensity distribution in the metal at the dipole and quadrupole resonance frequencies, illustrating the expected dipole and quadrupole nature of the resonances. Above the plasma energy $\hbar \omega_\text{p}$ we see the characteristic additional resonances in the nonlocal model due to the excitation of longitudinal modes, as previously reported for different metal nanoparticles \cite{Ruppin:1973a,Raza:2011a,Fuchs:1987a}.

The difference between the red and green curves in Fig. \ref{fig:fig1} shows the importance of taking into account interband transitions in the response of the metal shell. The implications on the dipole and quadrupole resonances are that they are redshifted and damped due to interband transitions, with greatest impact on the quadrupole peak. In the remaining part of this paper, we will always use measured values for the dielectric function \cite{Rakic:1998a}, i.e. we take interband transitions into account. We will concentrate on the dipole resonance, since this peak is the strongest, is close to visible and infrared frequencies and can be affected by the shape and size of the cylinder and the background permittivity.

\begin{figure*}[!tb]
\center
\includegraphics[width=0.85\textwidth]{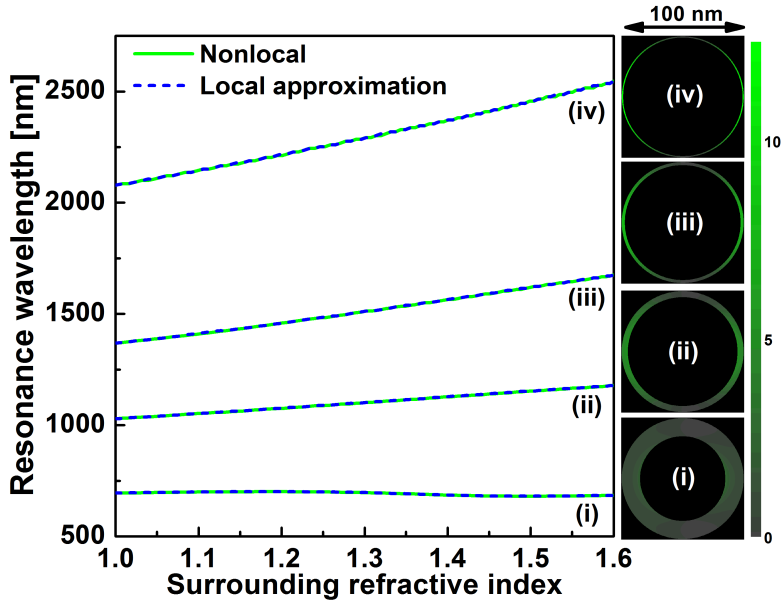}
\caption{The dipole resonance wavelength calculated with both local and nonlocal response taking into account interband transitions as a function of the background refractive index for four different $r_1/r_2$ ratios: (i) 0.7 (ii) 0.9 (iii) 0.95 (iv) 0.98. The outer radius of the nanotube is kept constant at $r_2 = 100\,\text{nm}$. The panel on the right shows the normalized intensity distribution $|\mathbf E|^2/|\mathbf{E_0}|^2$ in the nonlocal model at the vacuum dipole resonance wavelength for the corresponding four different shapes.}
\label{fig:fig2}
\end{figure*}

There are two geometrical properties that can be modified in the nanotube structure: the first is the shape defined by the $r_1/r_2$ ratio and the second is the overall size, that is, varying the outer radius $r_2$ but keeping $r_1/r_2$ constant. In Fig. \ref{fig:fig2} we show the effect of shape variations of the nanotube on its sensing abilities, which is quantified through the change in the dipole resonance wavelength when the background refractive index is increased. We see that regardless of the shape, the dependency is always approximately linear. However, as shown in Fig. \ref{fig:fig2}(i) there is no significant dependency on the background refractive index for low $r_1/r_2$ ratios, indicating the lack of ability to sense. Only when the shell becomes thin ($r_1/r_2 \rightarrow 1$) does the resonance wavelength shift with the refractive index. The thinner the shell, the greater is the average slope of the curves. Relaxing the nonlocal description to a local one does not change this trend, because the dipole resonances occur at too low energies for the nonlocal blueshift to kick in. Furthermore, the resonance wavelength shifts to higher wavelengths when the shell becomes thinner, because the coupling between the cavity and cylinder modes increases. Thus, even though Fig. \ref{fig:fig2}(iv) represents a nanotube with a $2\,\text{nm}$ thin metal shell, where nonlocal blueshifts are expected to be very prominent, the local approximation predicts sensitivities that are almost identical to the nonlocal description. So, as in Fig. \ref{fig:fig1}, here in Fig. \ref{fig:fig2} we see that for ultra-thin nanotubes the usual observation of larger nonlocal blueshifts for smaller structures does not occur. The nonlocal blueshift cancels out with the decrease of the resonance energy due to increased hybridization.

\begin{table}[b]
\center
\caption{Sensitivity and figure-of-merit calculations Eq.~(\ref{eq:FOM}) in the nonlocal description at the refractive index of air $n_\text{b} = 1$ (for gas sensing) and water $n_\text{b}= 1.333$ (for liquid sensing) for the four different shapes of Fig. \ref{fig:fig2}.}
\begin{tabular}{ccccc}
  \toprule
  \multicolumn{1}{c}{\multirow{2}{*}{$(r_1,r_2)$}} & \multicolumn{2}{c}{$\partial \lambda/\partial\text{n}_\text{b}$ [nm/RIU]} & \multicolumn{2}{c}{FOM} \\ \cmidrule(r){2-3} \cmidrule(r){4-5}
  & $n_\text{b} = 1$ & $n_\text{b}= 1.333$ & $n_\text{b} = 1$ & $n_\text{b}= 1.333$ \\ \midrule
  $(70\,\text{nm},100\,\text{nm})$ & 58 & -103 & 0.3 & 0.4 \\
  $(90\,\text{nm},100\,\text{nm})$ & 298 & 261 & 1.6 & 1.2 \\
  $(95\,\text{nm},100\,\text{nm})$ & 470 & 539 & 1.9 & 1.9 \\
  $(98\,\text{nm},100\,\text{nm})$ & 790 & 788 & 2.4 & 2.2 \\
  \toprule
\end{tabular}
\label{tab:tab1}
\end{table}

For a more quantitative description of the sensitivity of the nanotube, we present sensitivity and FOM calculations of the nanotube structures shown in Fig.~ \ref{fig:fig2} at the refractive index of air and water in Table~\ref{tab:tab1}. As in Fig.~\ref{fig:fig2}, it is again clear from Table~\ref{tab:tab1} that increased sensitivity can be achieved for thinner metal shells. Comparing the sensitivity of the nanotube with other LSPR sensors based on different nanoparticle geometries \cite{Mayer:2011a}, where the sensitivity is in the range $90-801\,$nm per refractive index unit (RIU), shows that the nanotube is comparable in sensitivity for ratios $r_1/r_2>0.7$, while it is superior for very high $r_1/r_2$ ratios. Comparison of the FOM with other nanoparticle LSP sensors also shows equally good performance by the nanotube, although the FOM is mainly dependent on the properties of Au and not easily improved by changing the geometry~\cite{Jeppesen:2010a}. The sensitivity values in Table~\ref{tab:tab1} also reveal that the nanotube has a high sensitivity at both the refractive index of air and water, which shows the versatility of a nanotube-based sensor and its applicability as both a gas and liquid sensor.

\begin{figure*}[tb]
\center
\includegraphics[width=0.85\textwidth]{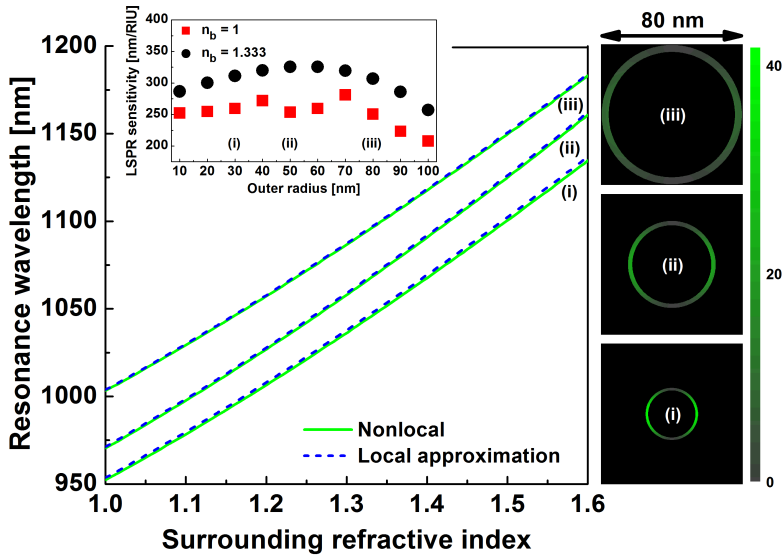}
\caption{The dipole resonance wavelength calculated with both local and nonlocal response taking into account interband transitions as a function of the background refractive index for three different $r_2$ values: (i) 30\,nm (ii) 50\,nm (iii) 80\,nm. The shape of the nanotube is kept constant by setting $r_1/r_2 = 0.9$. The panel on the right shows the normalized intensity distribution $|\mathbf E|^2/|\mathbf{E_0}|^2$ in the nonlocal model at the vacuum resonance wavelength for the corresponding three different sizes. Inset: The LSPR sensitivity at the refractive index of air ($n_\text{b}=1$) and water ($n_\text{b}=1.333$) calculated with the nonlocal model as a function of outer radius while keeping $r_1/r_2=0.9$.}
\label{fig:fig3}
\end{figure*}

Besides shape variations, we also varied the size $r_2$ of the nanotube, while keeping $r_1/r_2$ constant. Figure \ref{fig:fig3} depicts the dipole resonance wavelength as a function of the background refractive index for three different sizes with $r_1/r_2=0.9$. The sensing ability of the nanotube is not as dependent on size as it is on shape, which can be seen by the three almost parallel lines in Fig. \ref{fig:fig3}. Even though the sensitivity does not change much with increasing size, there is still an optimum size which occurs at $r_2=50\,\text{nm}$ and $r_2=70\,\text{nm}$ for liquid and gas sensing, respectively, see the inset of Fig. \ref{fig:fig3}. The fact that it is neither the smallest nor the biggest nanotube size that gives the highest sensitivity can be explained by a trade-off between the total structure size and the shell thickness. If the size of the structure is too small, then we have a weak LSP excitation and thereby poor sensing ability, but if the structure size is too big (with the shape kept constant) the absolute shell thickness increases, which also decreases the sensing ability, as we saw in Fig. \ref{fig:fig2}. Therefore, for a larger $r_1/r_2$ value the optimum size will also be larger.

In Fig.~\ref{fig:fig3} we also show the calculations using the local approximation. As seen, effects are surprisingly well accounted for even with a local description, despite the fact that we actually consider very thin metallic shells, for instance a 3\,nm shell in Fig. \ref{fig:fig3}(iii), with concomitant strong plasmon hybridization. The strong hybridization in ultra-thin metal shells shifts the dipole resonance to very low energies, where the nonlocal blueshift is weak. The sensitivity and consequently the FOM are therefore weakly influenced by nonlocal response. Although it is hardly visible in Fig. \ref{fig:fig3}, the local resonances do in fact occur at slightly longer wavelengths than in the nonlocal description, revealing a small nonlocal blueshift.

\section{Conclusions and outlook} \label{sec:con}
We have examined the infinite single dielectric-metal nanotube structure as an approximation for a dilute array of nanotubes with high aspect ratio. We calculate the extinction properties of a silica-gold nanotube analytically for both local and nonlocal response by extending the Mie theory for nanowires to nanotube geometries. Our investigation reveals that in contrast to the spherical nanoshell~\cite{Tam:2004a}, the sensing ability of the nanotube is highly dependent on the shape of the structure, where few-nm thin shells produce extreme sensitivities. The sensitivity is shown to be less dependent on the overall structure size. The sensitivity at the refractive index of air and water of ultra-thin nanotubes are superior to other nanoparticle geometries, making nanotubes very promising for both gas and liquid sensing.

Our results also show unexpectedly that nonlocal response has negligible influence on the extinction and sensing properties of the nanotube, even though the metal shell is ultra-thin (a few nm), because the hybridization in the nanotube is so strong that the dipole resonance is pushed to very low energies. The strength of the nonlocal blueshift is an interplay between the metal thickness and the resonance energy, where a thinner shell produces a stronger blueshift while a lower energy produces a weaker blueshift. This interplay is surprisingly well-balanced in the nanotube structure, because a thinner shell gives rise to lower resonance energies.

With the high sensitivity and good FOM of the nanotube geometry, we propose a sensor based on ultra-thin nanotubes. The robustness of the sensitivity of the nanotube to size variations provides desirable advantages, since fluctuations in size due to imperfect fabrication will have less impact. In the special case of gas sensing, the sensitivity may be further improved by a factor of two by designing the nanotube to have a hollow core. With a hollow core, the inner surface of the metal shell is also exposed to the surrounding medium, which significantly improves the sensitivity. However, mechanical stability is sacrificed with a hollow core if for instance the nanotubes are to stand vertically on a substrate.

\section{Appendix} \label{sec:app}
The nonlocal optical properties of the nanotube are determined by solving Maxwell's wave equation coupled to the hydrodynamic equation for the current~\cite{Raza:2011a}. We solve the coupled set of equations by extending the Mie theory for wires of Ref.~\cite{Ruppin:2001a} to core-shell structures. By expanding the electromagnetic fields in the dielectric core, metal shell and surrounding medium in cylindrical Bessel functions, we can most easily take into account Maxwell's boundary conditions along with the additional boundary condition of a vanishing normal component of the current in the nonlocal case~\cite{Raza:2011a}. Although quantum tunneling is not taken into account with this treatment, we do not expect any such effects to be important in this structure \cite{Zuloaga:2009a,Ozturk:2011a}.

To determine the extinction property of the infinite cylindrical nanotube we calculate the extinction cross section~\cite{HulstBook}
\begin{equation}
    \sigma_\text{ext} = -\frac{2}{k_0r_2} \sum_{n=-\infty}^{\infty} {\rm Re}\{a_n\}, \label{eq:sigma}
\end{equation}
where $k_0=\sqrt{\epsilon_\text{b}} \omega/c$ is the background wave vector, $\epsilon_\text{b}$ is the background permittivity and $a_n$ is a cylindrical Bessel-function expansion coefficient for the scattered electromagnetic field. We consider a normally incident electric-field polarization perpendicular to the cylinder axis (TM), as sketched in the inset of Fig. \ref{fig:fig1}. The nonlocal-response scattering coefficient is calculated analytically as
\begin{equation}
    a_n=-\frac{\sqrt{\epsilon_\text{b}} J_n(k_0 r_2) \left[C_n + J_n^\prime P_n - H_n^\prime Q_n\right] - \sqrt{\epsilon} J_n^\prime(k_0 r_2) \left[J_n P_n - H_n Q_n\right]} {\sqrt{\epsilon_\text{b}} H_n(k_0 r_2) \left[C_n+ J_n^\prime P_n - H_n^\prime Q_n\right] - \sqrt{\epsilon} H_n^\prime(k_0 r_2) \left[J_n P_n - H_n Q_n\right]}. \label{eq:anNL}
\end{equation}
Here, $J_n$ and $H_n$ are the Bessel and Hankel functions of the first kind, $k_\text{t} = \sqrt{\epsilon} \omega/c$ and $\epsilon(\omega)=\epsilon_\text{other}(\omega) - \omega_\text{p}^2/(\omega[\omega+i\gamma])$ is the Drude local-response function that includes interband effects through $\epsilon_\text{other}(\omega)$. The argument of the Bessel and Hankel functions are $k_\text{t} r_2$ unless written explicitly otherwise.

The coefficients $P_n$, $Q_n$ and $C_n$ are given by
\begin{align}
    P_n &= p_n \alpha_n + J_n(k_\text{c}r_1) \left[H_n(k_\text{t}r_1) \delta_n + H_n \tau_n \right], \label{eq:Pn} \\
    Q_n &= q_n \alpha_n + J_n(k_\text{c}r_1) \left[J_n(k_\text{t}r_1) \delta_n + J_n \tau_n \right], \label{eq:Qn} \\
    C_n &= \frac{i n}{k_0 r_2} \left[ H_n(k_\text{l}r_2) c_n - J_n(k_\text{l}r_2) d_n \right], \label{eq:Cn}
\end{align}
 where $k_\text{c} = \sqrt{\epsilon_\text{c}} \omega/c$ and $\epsilon_\text{c}$ is the dielectric constant of the core. Furthermore, $k_\text{l}^2=(\omega^2+i\omega\gamma-\omega_\text{p}^2/\epsilon_\text{other})/\beta^2$ and $\beta^2=3 v_\text{F}^2/5$ with $v_\text{F}$ being the Fermi velocity of the metal shell. The coefficients $p_n$, $q_n$, $\alpha_n$, $\delta_n$ and $\tau_n$ of Eqs.~(\ref{eq:Pn}-\ref{eq:Qn}) are given as
\begin{align}
    p_n &= \sqrt{\epsilon} J_n^\prime(k_\text{c} r_1) H_n(k_\text{t} r_1) - \sqrt{\epsilon_\text{c}} J_n(k_\text{c} r_1) H_n^\prime(k_\text{t} r_1), \label{eq:pn} \\
    q_n &= \sqrt{\epsilon} J_n^\prime(k_\text{c} r_1) J_n(k_\text{t} r_1) - \sqrt{\epsilon_\text{c}} J_n(k_\text{c} r_1) J_n^\prime(k_\text{t} r_1). \label{eq:qn} \\
    \alpha_n &= \left(\frac{k_\text{l} \epsilon_\text{other}}{k_0}\right)^2 \left[J_n^\prime(k_\text{l}r_2) H_n^\prime(k_\text{l}r_1)-H_n^\prime(k_\text{l}r_2)J_n^\prime(k_\text{l}r_1) \right], \label{eq:alphan} \\
    \delta_n &= -\frac{k_\text{l}n^2 \sqrt{\epsilon_\text{c}} \epsilon_\text{other} (\epsilon-\epsilon_\text{other})}{k_\text{t}k_0^2 r_1^2} \left[ J_n^\prime(k_\text{l}r_2) H_n(k_\text{l}r_1)- H_n^\prime(k_\text{l}r_2) J_n(k_\text{l} r_1) \right], \label{eq:deltan} \\
    \tau_n &= -\frac{k_\text{l}n^2 \sqrt{\epsilon_\text{c}} \epsilon_\text{other} (\epsilon-\epsilon_\text{other})}{k_\text{t}k_0^2 r_1 r_2} \left[ H_n^\prime(k_\text{l}r_1) J_n(k_\text{l}r_1)- J_n^\prime(k_\text{l}r_1) H_n(k_\text{l} r_1) \right], \label{eq:taun}
\end{align}
while the coefficients $c_n$ and $d_n$ of Eq.~(\ref{eq:Cn}) are given as
\begin{align}
    c_n &=f_n \left[ J_n^\prime(k_\text{l}r_2) \eta_n + J_n(k_\text{l} r_1) \kappa_n \right] + J_n^\prime(k_\text{l}r_1) g_n \left[ J_n p_n - H_n q_n \right], \label{eq:cn} \\
    d_n &=f_n \left[ H_n^\prime(k_\text{l}r_2) \eta_n + H_n(k_\text{l} r_1) \kappa_n \right] + H_n^\prime(k_\text{l}r_1) g_n \left[ J_n p_n - H_n q_n \right], \label{eq:dn}
\end{align}
where
\begin{align}
    g_n &= \frac{ink_\text{l} \epsilon_\text{other}(\epsilon-\epsilon_\text{other})}{k_0 k_\text{t} r_2}, \label{eq:gn} \\
    f_n &= \frac{in\sqrt{\epsilon_\text{c}} (\epsilon-\epsilon_\text{other})}{k_0 k_\text{t} r_1} J_n(k_\text{t} r_1), \label{eq:fn} \\
    \eta_n &= k_\text{l} \left[ J_n(k_\text{t}r_1) H_n^\prime(k_\text{t}r_1) - H_n(k_\text{t} r_1) J_n^\prime(k_\text{t} r_1) \right], \label{eq:etan} \\
    \kappa_n &= \frac{n^2 (\epsilon-\epsilon_\text{other})}{k_\text{t} r_2 r_1} \left[ J_n(k_\text{t}r_1) H_n - H_n(k_\text{t}r_1) J_n \right]. \label{eq:kappan}
\end{align}
The local-response result can be retrieved in the limit of a vanishing Fermi velocity for which $P_n=p_n$, $Q_n=q_n$ and $C_n=0$.



\begin{thebibliography}{10}
\providecommand{\url}[1]{{#1}}
\providecommand{\urlprefix}{URL }
\expandafter\ifx\csname urlstyle\endcsname\relax
  \providecommand{\doi}[1]{DOI \discretionary{}{}{}#1}\else
  \providecommand{\doi}{DOI \discretionary{}{}{}\begingroup
  \urlstyle{rm}\Url}\fi

\bibitem{Anker:2008a}
J.N. Anker, W.P. Hall, O.~Lyandres, N.C. Shah, J.~Zhao, R.P. van Duyne, Nat.
  Mater. \textbf{7}, 442 (2008)

\bibitem{Prodan:2003a}
E.~Prodan, C.~Radloff, N.J. Halas, P.~Nordlander, Science \textbf{302}, 419
  (2003)

\bibitem{Brongersma:2003a}
M.L. Brongersma, Nat. Mater. \textbf{2}, 296 (2003)

\bibitem{Raschke:2004a}
G.~Raschke, S.~Brogl, A.S. Susha, A.L. Rogach, T.A. Klar, J.~Feldmann,
  B.~Fieres, N.~Petkov, T.~Bein, A.~Nichtl, K.~K\"{u}rzinger, Nano Lett.
  \textbf{4}, 1853 (2004)

\bibitem{Nehl:2004a}
C.L. Nehl, N.K. Grady, G.P. Goodrich, F.~Tam, N.J. Halas, J.H. Hafner, Nano
  Lett. \textbf{4}, 2355 (2004)

\bibitem{Tam:2004a}
F.~Tam, C.~Moran, N.J. Halas, J. Phys. Chem. B \textbf{108}, 17290 (2004)

\bibitem{Bardhan:2011a}
R.~Bardhan, S.~Lal, A.~Joshi, N.J. Halas, Acc. Chem. Res. \textbf{44}, 936
  (2011)

\bibitem{Prodan:2004a}
E.~Prodan, P.~Nordlander, J. Chem. Phys. \textbf{120}, 5444 (2004)

\bibitem{David:2011a}
C.~David, F.J. Garc\'{i}a~de Abajo, J. Phys. Chem. C \textbf{115}, 19470 (2011)

\bibitem{Toscano:2012a}
G.~Toscano, S.~Raza, A.-P. Jauho, N.A. Mortensen, M.~Wubs, Opt. Express
  \textbf{20}, 4176 (2012)

\bibitem{Boardman:1977a}
A.D. Boardman, B.V. Paranjape, J. Phys. F: Met. Phys. \textbf{7}, 1935 (1977)

\bibitem{Dasgupta:1981a}
B.B. Dasgupta, R.~Fuchs, Phys. Rev. B \textbf{24}, 554 (1981)

\bibitem{Ruppin:1973a}
R.~Ruppin, Phys. Rev. Lett. \textbf{31}, 1434 (1973)

\bibitem{Raza:2011a}
S.~Raza, G.~Toscano, A.-P. Jauho, M.~Wubs, N.A. Mortensen, Phys. Rev. B
  \textbf{84}, 121412(R) (2011)

\bibitem{Jones:1969a}
W.E. Jones, K.L. Kliewer, R.~Fuchs, Phys. Rev. \textbf{178}, 1201 (1969)

\bibitem{McPhillips:2010a}
J.~McPhillips, A.~Murphy, M.P. Jonsson, W.R. Hendren, R.~Atkinson,
  F.~H\"{o}\"{o}k, A.V. Zayats, R.J. Pollard, ACS Nano \textbf{4}, 2210 (2010)

\bibitem{Lim:2012a}
M.A. Lim, D.H. Kim, C.O. Park, Y.W. Lee, S.W. Han, Z.~Li, R.S. Williams,
  I.~Park, ACS Nano \textbf{6}, 598 (2012)

\bibitem{Schroter:2001a}
U.~Schr\"{o}ter, A.~Dereux, Phys. Rev. B \textbf{64}, 125420 (2001)

\bibitem{Zhu:2011a}
J.~Zhu, K.F. Li, Eur. Phys. J. B \textbf{80}, 83 (2011)

\bibitem{Rakic:1998a}
A.D. Raki\'{c}, A.B. Djuri\v{s}i\'{c}, J.M. Elazar, M.L. Majewski, Appl. Opt.
  \textbf{37}, 5271 (1998)

\bibitem{Fuchs:1987a}
R.~Fuchs, F.~Claro, Phys. Rev. B \textbf{35}, 3722 (1987)

\bibitem{Mayer:2011a}
K.M. Mayer, J.H. Hafner, Chem. Rev. \textbf{111}, 3828 (2011)

\bibitem{Jeppesen:2010a}
C.~Jeppesen, S.~Xiao, N.A. Mortensen, A.~Kristensen, Opt. Express \textbf{18},
  25075 (2010)

\bibitem{Ruppin:2001a}
R.~Ruppin, Opt. Commun. \textbf{190}, 205 (2001)

\bibitem{Zuloaga:2009a}
J.~Zuloaga, E.~Prodan, P.~Nordlander, Nano Lett. \textbf{9}, 887 (2009)

\bibitem{Ozturk:2011a}
Z.F. \"{O}zt\"{u}rk, S.~Xiao, M.~Yan, M.~Wubs, A.-P. Jauho, N.A. Mortensen, J.
  Nanophotonics \textbf{5}, 051602 (2011)

\bibitem{HulstBook}
H.~van~de Hulst, \emph{Light Scattering by Small Particles} (John Wiley \&
  Sons, Inc., New York, 1957)

\end{thebibliography}
\end{document}